# Penguin-like Diagrams from the Standard Model

Chia Swee Ping

*High Impact Research, University of Malaya, 50603 Kuala Lumpur, Malaysia*

**Abstract.** The Standard Model is highly successful in describing the interactions of leptons and quarks. There are, however, rare processes that involve higher order effects in electroweak interactions. One specific class of processes is the penguin-like diagram. Such class of diagrams involves the neutral change of quark flavours accompanied by the emission of a gluon (gluon penguin), a photon (photon penguin), a gluon and a photon (gluon-photon penguin), a Z-boson (Z penguin), or a Higgs-boson (Higgs penguin). Such diagrams do not arise at the tree level in the Standard Model. They are, however, induced by one-loop effects. In this paper, we present an exact calculation of the penguin diagram vertices in the 'tHooft-Feynman gauge. Renormalization of the vertex is effected by a prescription by Chia and Chong which gives an expression for the counter term identical to that obtained by employing Ward-Takahashi identity. The on-shell vertex functions for the penguin diagram vertices are obtained. The various penguin diagram vertex functions are related to one another via Ward-Takahashi identity. From these, a set of relations is obtained connecting the vertex form factors of various penguin diagrams. Explicit expressions for the gluon-photon penguin vertex form factors are obtained, and their contributions to the flavor changing processes estimated.

## INTRODUCTION

The Standard Model (SM) of elementary particles is based on the SU(3)×SU(2)×U(1) gauge group. With three families of leptons and quarks, the theory is highly successful in confronting experimental observations. The recent discovery of the Higgs boson [1, 2] has provided a great boost to the SM.

Flavour mixing in the quark sector with respect to the weak interactions is one of the salient features of the SM. With three families of quarks, the flavour mixing matrix, the Cabibbo-Kobayashi-Maskawa (CKM) matrix $V_{ij}$ [3, 4], contains a complex phase angle. This leads to the consequence that direct CP violation effects are predicted in a natural manner in such rare processes.

Neutral flavour-changing transition processes as exemplified by the 'penguin' diagrams are one of the salient features of the SM. Such processes do not occur at the tree level. It occurs, however, at the one-loop level which involves an internal W boson and a quark lines. The neutral change of flavour comes about because the coupling of W boson to the quark involves the CKM matrix, which mixes quarks from different generations. A penguin diagram therefore couples to quarks of different flavours, but of the same charge. The penguin diagram may be mediated by the gluon (gluon penguin) [5, 6], the photon (photon penguin), the $Z^0$ boson (Z penguin) [7-15], or the Higgs boson (Higgs penguin) [16-18]. The Higgs penguin differs from the other penguins in that the Higgs boson is scalar, whereas all others mediatory bosons are vector.

More complicated penguin diagrams with the internal emission of two neutral gauge bosons, such as the two-gluon penguin, the two-photon penguin and the gluon-photon penguin, may also be of interest [19-22].

In this paper, I shall present the calculation of the vertex functions for the following penguin diagrams: (a) gluon penguin, (b) photon penguin, and (c) gluon-photon penguin. The calculation is performed in the 'tHooft-Feynman gauge. Divergence encountered in the vertex function is eliminated by using a renormalization scheme proposed by Chia and Chong [13]. It is easily demonstrated that the counter term as calculated from the renormalization scheme above removes the divergence and yields a result identical to that would be obtained by using Ward-Takahashi Identity [23, 24].

## The Gluon Penguin

We first calculate the flavour-changing quark self energy [25], the Feynman diagram for which is depicted in Fig. 1. In the 'tHooft-Feynman gauge, this diagram is to be supplemented by another diagram in which the internal W-boson line is replaced by the unphysical charged Higgs line. This procedure of including the unphysical charged

Higgs line for each internal W-boson line is repeated in the subsequent sections of penguin diagrams in the calculation in the 'tHooft-Feynman gauge.

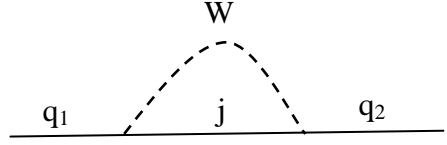

**FIGURE 1.** Feynman diagrams for the flavour-changing quark self energy.

The resultant self energy is divergent, and has to be renormalized using the scheme of Chia and Chong. The self energy $\Sigma(p)$, after renormalization, is given by

$$\Sigma(p) = -\frac{g^2}{16\pi^2 M_W^2}\sum_j \lambda_j F_j [(p^2 - m_2^2 - m_1^2)p.\gamma L + m_2 m_1(-p.\gamma R + m_2 R + m_1 L)] \tag{1}$$

where $\lambda_j = V_{j2}^* V_{j1}$ and $F_j$ is the self energy form factor given by

$$F_j = -\frac{x}{2(1-x)^4}\{(4-x)(1-x^2) + x(9 - 4x + x^2)\ln x\} \tag{2}$$

$$x = m_j^2/M_W^2 \tag{3}$$

It is noted that the flavour-changing self energy vanishes when one or both the external quark lines go on mass shell/
For the gluon penguin, the relevant Feynman diagram is shown in Fig. 2. The vertex function $\Gamma_\mu^a(p,q)$, after including diagram in which the internal W-boson line is replaced by m unphysical charged Higgs line and renormalization, is given by

$$\Gamma_\mu^a(p,q) = \frac{g^2 g_s \lambda^a/2}{16\pi^2 M_W^2}\sum_j \lambda_j\{E_1[-(m_2^2 + m_1^2)\gamma_\mu L - m_2 m_1 R + q^2\gamma_\mu L + 2(p^2 - p.q)\gamma_\mu L$$
$$+ 1/2(2p.\gamma\gamma_\mu\gamma.p - p.\gamma\gamma_\mu\gamma.q - q.\gamma\gamma_\mu\gamma.p)L] + 2E_2(q_\mu q.\gamma - q^2\gamma_\mu)L + E_3(p.\gamma\gamma_\mu\gamma.q$$
$$- q.\gamma\gamma_\mu\gamma.p)\} \tag{4}$$

where $E_1$, $E_2$, and $E_3$ are vertex form factors. These form factors are functions of $x = m_j^2/M_W^2$.

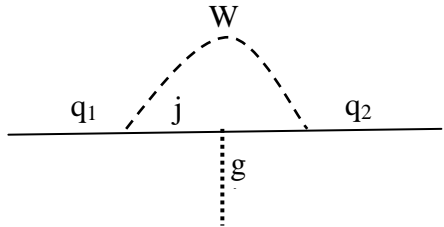

**FIGURE 2.** Feynman diagrams for the gluon penguin.

## The Photon Penguin

The relevant Feynman diagram for the gluon penguin is shown in Fig. 3. The renormalized vertex function $\Gamma_\nu(p,k)$ is given by

$$\Gamma_\nu(p,k) = \frac{eg^2}{16\pi^2 M_W^2} \sum_j \lambda_j \{\tilde{E}_1[-(m_2^2 + m_1^2)\gamma_\nu L - m_2 m_1 \gamma_\nu R + k^2 \gamma_\nu L + 2(p^2 - p.k)\gamma_\nu L$$
$$+ 1/2(2p.\gamma\gamma_\nu\gamma.p - p.\gamma\gamma_\nu\gamma.k - k.\gamma\gamma_\nu\gamma.p)L] + 2\tilde{E}_2(k_\nu k.\gamma - k^2\gamma_\nu)L + \tilde{E}_3(p.\gamma\gamma_\nu\gamma.k$$
$$- k.\gamma\gamma_\nu\gamma.p)\} \quad (5)$$

where $\tilde{E}_1, \tilde{E}_2$ and $\tilde{E}_3$ are vertex form factors for the photon penguin, which are functions of $x = m_j^2/M_W^2$.

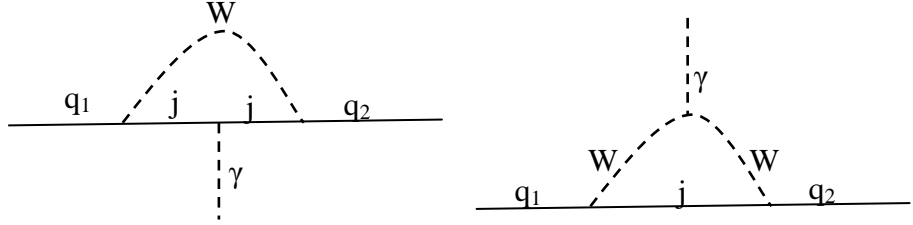

**FIGURE 3.** Feynman diagrams for the photon penguin.

## The Gluon-Photon Penguin

Fig. 4 depicts the gluon-photon penguin. This penguin involves the emission of two gauge bosons, a gluon and a photon. The vertex function as calculated in the 't Hooft-Feynman gauge is finite. Therefore, no renormalization is needed. In the calculation, we have assumed that the external quark masses and all external momenta are small compared to $M_W$. Retaining only terms linear in external momenta, the vertex function is given by

$$\Lambda_{\mu\nu}^a(p,q,k) = \frac{eg^2 g_s \lambda^a/2}{16\pi^2 M_W^2} \sum_j \lambda_j R\{D_1 (2p - q - k)^\sigma (\gamma_\sigma g_{\mu\nu} + \gamma_\mu g_{\nu\sigma} + \gamma_\nu g_{\mu\sigma}) - i\varepsilon_{\mu\nu\rho\sigma}\gamma^\sigma(D_2 q^\rho + D_3 k^\rho)\} \quad (6)$$

where $D_1$, $D_2$ and $D_3$ are the form factors associated with the gluon-photon penguin vertex function.

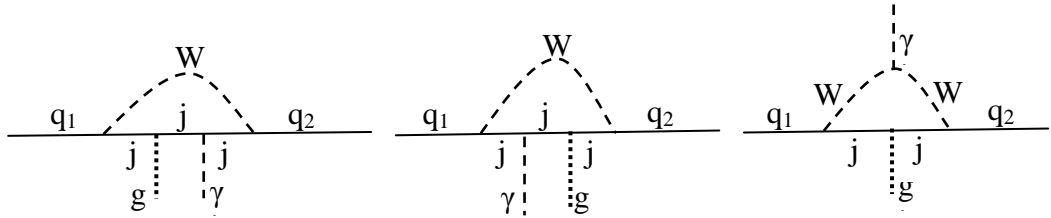

**FIGURE 4.** Feynman diagrams for the gluon-photon penguin.

## Ward-Takahashi Identity

The various vertex functions in the preceding sections are related via Ward-Takahashi Identity. The gluon penguin vertex function $\Gamma_\mu^a(p,q)$, for example, is related to the flavour changing self energy by

$$q^\mu \Gamma_\mu^a(p,q) = g_s \lambda^a/2 [\Sigma(p) - \Sigma(p-q)] \quad (7)$$

This gives the following relation

$$E_1 = F_j \quad (8)$$

Similarly, the photon penguin vertex function $\Gamma_\nu(p,k)$ is related to the flavour changing self energy by

$$k^\nu \Gamma_\nu(p,k) = -\frac{1}{3}e[\Sigma(p) - \Sigma(p-k)] \tag{9}$$

which results in the following relation

$$\tilde{E}_1 = -\frac{1}{3}F_j \tag{10}$$

On the other hand, the gluon- photon penguin vertex function is related to the

$$k^\nu \Lambda^a_{\mu\nu}(p,q,k) = -\frac{1}{3}e[\Gamma^a_\mu(p,q) - \Gamma^a_\mu(p-k,q)] \tag{11}$$

$$q^\mu \Lambda^a_{\mu\nu}(p,q,k) = g_s \frac{\lambda^a}{2}[\Gamma_\nu(p,k) - \Gamma_\nu(p-q,k)] \tag{12}$$

These give rise to the following relations

$$D_1 = \tilde{E}_1 = -\frac{1}{3}E_1 \tag{13}$$

$$D_2 = 2\tilde{E}_3 \tag{14}$$

$$D_3 = \frac{2}{3}E_3 \tag{15}$$

## Contribution of the Gluon-Photon Penguin to Flavour Changing Processes

The form factors associated with the gluon-photon penguin vertex function are given by

$$D_1 = \frac{1}{6}\{x(1-x^2)(4-x) + x^2(9-4x+x_2)lnx\}/(1-x)^4 \tag{16}$$

$$D_2 = \frac{1}{6}\{x(1-x)(11+9x) + x^2(21-x)lnx\}/(1-x)^3 \tag{17}$$

$$D_3 = -\frac{1}{6}\{2x(1-x) + x^2(3-x)lnx\}/(1-x)^3 \tag{18}$$

Table 1 displays the numerical values of $D_1$, $D_2$, and $D_3$ as functions of $x_j$.

**TABLE (1).** Numerical values of $D_1$, $D_2$, and $D_3$ as functions of $x_j$.

| j | $m_j$ (GeV) | $x_j$ | $D_1$ | $D_2$ | $D_3$ |
|---|---|---|---|---|---|
| u | 0.0023 | 0.8187E-09 | 0.5458E-09 | −0.2729E-09 | 0.1501E-08 |
| c | 1.275 | 0.2516E-03 | 0.1671E-03 | −0.8364E-04 | 0.4597E-03 |
| t | 173.07 | 0.4635E+01 | 0.4332E+00 | −0.3039E+00 | 0.1211E+01 |

We will consider the contribution of the gluon-photon penguin to the following flavour changing transitions:
(a) $s \rightarrow d$
(b) $b \rightarrow d$
(c) $b \rightarrow s$

To do this, we need a good knowledge of the CKM matrix elements, which have the following magnitudes [26]:

$$V = \begin{pmatrix} 0.97427 & 0.22534 & 0.00351 \\ 0.22520 & 0.97344 & 0.0412 \\ 0.00867 & 0.0404 & 0.999146 \end{pmatrix} \qquad (19)$$

The phases of the CKM elements are not well determined. However, we take advantage of the unitarity triangle relation offered by

$$\lambda_u + \lambda_c + \lambda_t = 0 \qquad (20)$$

to have a good estimate of the relative phases among the three $\lambda_j$. Table 2 shows the values of $\lambda_j$ for three transitions considered.

**TABLE (2).** Values of $\lambda_j$ for the three flavour changing transitions considered.

| Transition | $\lambda_u$ | $\lambda_c$ | $\lambda_t$ |
|---|---|---|---|
| $s \to d$ | 0.21954 | $-0.21919$ | $-0.32336 \times 10^{-3}$ |
|  | -- | $+ i\, 0.13570 \times 10^{-3}$ | $- i\, 0.13570 \times 10^{-3}$ |
| $b \to d$ | $-0.12254 \times 10^{-2}$ | $0.92782 \times 10^{-2}$ | $-0.80528 \times 10^{-2}$ |
|  | $- i\, 0.31926 \times 10^{-2}$ | -- | $+ i\, 0.31926 \times 10^{-2}$ |
| $b \to s$ | $0.27119 \times 10^{-3}$ | $-0.40099 \times 10^{-1}$ | $0.40365 \times 10^{-1}$ |
|  | $- i\, 0.74462 \times 10^{-3}$ | $+ i\, 0.74462 \times 10^{-3}$ | -- |

We now proceed to calculate the combined effects of $\lambda_j$ ($j = u, c, b$) and gluon-photon penguin vertex form factors $D_i$ (i = 1, 2, 3) for each of the three transitions in Table 2. These values are tabulated in Table 3. It is to be noted that for each of the transitions there is an arbitrary phase in the CKM matrix elements. We take the liberty to fix this arbitrary phase so that the $j = u$ contribution in $s \to d$ transition, he $j = c$ contribution in $b \to d$ transition, and he $j = t$ contribution in $b \to s$ transition are all real.

It is apparent from Table 3 that for $j = u$, the contribution is small for all the three transitions. Both $j = c$ and $j = t$ contribute in $s \to d$ transition to the magnitude of order $10^{-4}$. The $j = c$ contribution is mainly real, whereas the $j = t$ contribution has real and imaginary parts of equal strength. On the other hand $j = t$ is the dominant contribution in both $b \to d$ and $b \to s$ transitions with magnitude of order $10^{-3}$ and $10^{-2}$ respectively. The relative phases of the CKM matrix elements are chosen so that the $j = t$ contribution in the $b \to s$ transitions is purely real. With this choice of phases, the $j = t$ contribution in $b \to d$ transition has equal amount of real and imaginary parts.

**TABLE (3).** Values of combined contributions $\lambda_j D_i$ calculated from Tables 1 and 2.

| Transition | $j$ | $\lambda_j D_1(x_j)$ | $\lambda_j D_2(x_j)$ | $\lambda_j D_3(x_j)$ |
|---|---|---|---|---|
| $s \to d$ | $u$ | $1.198 \times 10^{-10}$ | $-5.991 \times 10^{-11}$ | $3.295 \times 10^{-10}$ |
|  | $c$ | $-3.663 \times 10^{-5}$ | $1.833 \times 10^{-5}$ | $-1.008 \times 10^{-4}$ |
|  |  | $+ i\, 2.268 \times 10^{-8}$ | $- i\, 1.135 \times 10^{-8}$ | $+ i\, 6.238 \times 10^{-8}$ |
|  | $t$ | $-1.401 \times 10^{-4}$ | $9.827 \times 10^{-5}$ | $-3.916 \times 10^{-4}$ |
|  |  | $- i\, 5.879 \times 10^{-5}$ | $+ i\, 4.124 \times 10^{-5}$ | $- i\, 1.643 \times 10^{-4}$ |
| $b \to d$ | $u$ | $-6.688 \times 10^{-13}$ | $3.344 \times 10^{-13}$ | $-1.839 \times 10^{-12}$ |
|  |  | $- i\, 1.743 \times 10^{-12}$ | $+ i\, 8.713 \times 10^{-13}$ | $- i\, 4.792 \times 10^{-12}$ |
|  | $c$ | $1.550 \times 10^{-6}$ | $-7.760 \times 10^{-7}$ | $4.265 \times 10^{-6}$ |
|  | $t$ | $-3.488 \times 10^{-3}$ | $2.447 \times 10^{-3}$ | $-9.752 \times 10^{-3}$ |
|  |  | $+ i\, 1.383 \times 10^{-3}$ | $- i\, 9.702 \times 10^{-4}$ | $+ i\, 3.866 \times 10^{-3}$ |
| $b \to s$ | $u$ | $1.480 \times 10^{-13}$ | $-7.401 \times 10^{-14}$ | $4.071 \times 10^{-13}$ |
|  |  | $- i\, 4.064 \times 10^{-13}$ | $+ i\, 2.032 \times 10^{-13}$ | $- i\, 1.118 \times 10^{-12}$ |
|  | $c$ | $-6.701 \times 10^{-6}$ | $3.354 \times 10^{-6}$ | $-1.843 \times 10^{-5}$ |
|  |  | $+ i\, 1.244 \times 10^{-7}$ | $- i\, 6.228 \times 10^{-8}$ | $+ i\, 3.423 \times 10^{-7}$ |
|  | $t$ | $1.749 \times 10^{-2}$ | $-1.227 \times 10^{-2}$ | $4.888 \times 10^{-2}$ |

The three internal quark contributions for each of the transitions can now be added together. Thus we arrive at the following expression for the combined form factors:

$$\mathcal{H}_i = \sum_j \lambda_j D_i(x_j) \tag{21}$$

In terms of the combined form factors $\mathcal{H}_i$, Eq. (6) can then be expressed as

$$\Lambda^a_{\mu\nu}(p,q,k) = \frac{eg^2 g_s \lambda^{a/2}}{16\pi^2 M_W^2} R\{\mathcal{H}_1(2p-q-k)^\sigma(\gamma_\sigma g_{\mu\nu} + \gamma_\mu g_{\nu\sigma} + \gamma_\nu g_{\mu\sigma}) - i\varepsilon_{\mu\nu\rho\sigma}\gamma^\sigma(\mathcal{H}_2 q^\rho + \mathcal{H}_3 k^\rho)\} \tag{22}$$

The numerical values of the combined form factors $\mathcal{H}_i$ are as shown in Table 4. For the $s \rightarrow d$ transition, the combined form factors $\mathcal{H}_i$ have equal real and imaginary parts, and are of magnitude of order $10^{-4}$. For the $b \rightarrow d$ transition, $\mathcal{H}_i$ also have equal real and imaginary parts, but are of magnitude of order $10^{-3}$ or $10^{-2}$. The $b \rightarrow s$ transition, on the other hand, has almost purely real $\mathcal{H}_i$, with magnitude of order $10^{-2}$.

**TABLE (4). Values of** the combined form factors $\mathcal{H}_i$.

| Transition | $\mathcal{H}_1$ | $\mathcal{H}_2$ | $\mathcal{H}_3$ |
|---|---|---|---|
| $s \rightarrow d$ | $-1.198 \times 10^{-4}$ $- i\, 5.877 \times 10^{-5}$ | $1.166 \times 10^{-4}$ $+ i\, 4.123 \times 10^{-5}$ | $-4.924 \times 10^{-4}$ $- i\, 1.642 \times 10^{-4}$ |
| $b \rightarrow d$ | $-3.486 \times 10^{-3}$ $+ i\, 1.383 \times 10^{-3}$ | $2.446 \times 10^{-3}$ $- i\, 9.702 \times 10^{-4}$ | $-9.748 \times 10^{-3}$ $+ i\, 3.866 \times 10^{-3}$ |
| $b \rightarrow s$ | $1.748 \times 10^{-2}$ $+ i\, 1.244 \times 10^{-7}$ | $-1.227 \times 10^{-2}$ $- i\, 6.228 \times 10^{-8}$ | $4.886 \times 10^{-2}$ $+ i\, 3.423 \times 10^{-7}$ |

## CONCLUSION

We have presented the calculation of vertex functions for the gluon penguin, the photon penguin and the gluon-photon penguin within the framework of the Standard Model. Ward-Takahashi identity is invoked to relate the vertex functions of different penguin diagrams. A set of relations is then obtained among the vertex form factors of different penguin diagrams: Eqs. (8), (10), (13), (14) and (15).

For the gluon-photon penguin vertex the external momenta are assumed to be small compared W-boson mass. The vertex function can then be described in terms of the three vertex form factors $D_1$, $D_2$ and $D_3$, which are functions of $x = m_j^2/M_W^2$. Their numerical values are calculated, and their contributions to the flavor changing processes estimated.